\documentclass[%
 reprint,
 amsmath,amssymb,
aps,
pra,
longbibliography
]{revtex4-1}

\usepackage{graphicx}
\usepackage{dcolumn}
\usepackage{bm}

\usepackage{hyperref}

\begin{document}


\title{Long term measurement of the $^{87}$Sr clock frequency at the limit of primary Cs clocks}

\author{R. Schwarz, S. D\"orscher, A. Al-Masoudi, E. Benkler, T. Legero, U. Sterr, S. Weyers, J. Rahm, B. Lipphardt, C. Lisdat}

 \affiliation{Physikalisch-Technische Bundesanstalt, Bundesallee 100, 38116 Braunschweig, Germany}
 \email{christian.lisdat@ptb.de}
%




\date{\today}
%


\begin{abstract}
We report on a series of 42 measurements of the transition frequency of the 429~THz (5s$^2$)~$^1$S$_0$--(5s5p)~$^3$P$_0$ line in $^{87}$Sr taken over three years from 2017 to 2019. 
They have been performed at the Physikalisch-Technische Bundesanstalt (PTB) between the laboratory strontium lattice clock and the primary caesium fountain clocks CSF1 and CSF2. 
The length of each individual measurement run has been extended by use of a hydrogen maser as flywheel to improve the statistical uncertainty given by	 the Cs clocks. 
We determine an averaged transition frequency of $429\:228\:004\:229\:873.00(0.07)$~Hz with $1.5\times10^{-16}$ fractional uncertainty, at the limit of the current realization of the unit hertz.	 
Analysis of the data provides an improved limit on the coupling of the gravitational potential of the Sun to the proton--electron mass ratio $\mu$, and confirms the limits on its temporal drift.
\end{abstract}

\maketitle

%
\section{Introduction}
\label{Sec:intro}
%

The revision of the International System of Units (SI) in 2019 marked a fundamental evolution in metrology.  Several base units were redefined by	 assigning fixed values to fundamental constants \cite{bip19,goe19}. While it is likely that no further redefinitions will be needed in the future to keep up with scientific progress for these units, the same is not the case for the most precisely realized unit so far, the SI unit `second' \cite{biz19}.
Redefining the second by  fixing another fundamental constant (e.g. the electron mass $m_\mathrm{e}$) is not constructive, because ab initio calculations of atomic transition frequencies from fundamental constants are by far not competitive  to the accuracy of present realizations.
As progress in Cs clocks is saturating and the reduction of the uncertainties of optical clocks -- now in the low $10^{-18}$ range -- is ongoing, a redefinition using optical atomic transitions  is under discussion.
The Consultative Committee for Time and Frequency (CCTF) of the Bureau International des Poids et Mesures (BIPM) issued a roadmap, defining requirements that optical clocks have to fulfill before a redefinition can take place \cite{rie18}.
Besides comparisons of optical clocks at the level of $10^{-18}$ and contributions of optical clocks to International Atomic Time (TAI), this roadmap also requires three independent measurements of optical clocks to independent Cs primary clocks with fractional uncertainty below $3\times10^{-16}$. 

Even more fundamental, the base of all current precision measurements and the current SI is the assumption that fundamental constants are independent of space and time.
In this respect repeated measurements of the frequency ratios of clocks allow for tests of fundamental aspects of physics, e.g., they probe for variations of the fine-structure constant $\alpha$, the proton--electron mass ratio $\mu = m_\mathrm{p} / m_\mathrm{e}$ or the light quark mass to quantum chromodynamics mass scale ratio $X_\mathrm{q} = m_\mathrm{q}/\Lambda_\mathrm{QCD}$ over time or as function of position in a gravitational potential \cite{fla07c}. 

As optical frequency ratio measurements are mostly sensitive to variations of $\alpha$, comparisons between optical clocks and clocks based on hyperfine transitions, like in absolute frequency measurements, play an important role for investigations concerning $\mu$. 
Additionally, certain molecular transitions can be highly sensitive to variations of $\mu$ due to a cancellation between electronic and rovibrational energies \cite{fla07b,bag13,kob19}. 
Analysis of optical frequency ratios and absolute frequency measurements has therefore steadily lowered limits on the respective coupling coefficients and temporal drifts \cite{ros08, bla08, gue12, pei13, let13, god14, hun14, ash18, mcg19}.
These investigations provide laboratory-based complements to astrophysical investigations on molecules \cite{uba16,uba18}.

In this article, we analyze 42 frequency measurements of the 429~THz transition (5s$^2$)~$^1$S$_0$--(5s5p)~$^3$P$_0$ in $^{87}$Sr against PTB's primary caesium fountain clocks CSF1 and CSF2 \cite{wey18} accumulated over a period of three years. 
The data provide an improved value of the mean Sr transition frequency with $1.5\times10^{-16}$ fractional uncertainty, the lowest one for an absolute frequency so far, fulfilling one of the requirements for a revision of the SI second.
To exploit the performance of the Sr lattice clock with improved instability of $\sigma_{y, {\rm Sr}}=5 \times 10^{-17}/\sqrt{\tau/\mathrm{s}}$, we employ a hydrogen maser as flywheel oscillator to extend the 59~days total up-time of the Sr clock to about 120~days of comparison to the Cs clocks. 
 
As a test of position invariance, the data were analyzed for annual variations due to the varying gravitational potential on the Earth's elliptic orbit around the Sun. The amplitude of the corresponding fractional frequency variations amounts to $-18(86)\times 10^{-18}$.	
Disentangling the different possible contributions of varying fundamental constants using other measurements, we improve the limit on the coupling of the proton--electron mass ratio $\mu$ to a gravitational potential by a factor of 1.6 and confirm limits on its time dependence  \cite{mcg19}.

%
\section{Frequency standards}
\label{Sec:clocks}
%
The measurements described in this article involve three frequency standards, the Sr lattice clock and the two primary
caesium fountain clocks CSF1 and CSF2. 
Active hydrogen masers serve as flywheel oscillators to bridge the downtimes of the optical frequency standard.

The Sr standard is located in a different building than the microwave frequency standards. The buildings are linked by path-length stabilized optical fibers and high-quality radio-frequency (rf) cables. The frequency conversion between optical and microwave domain can be performed
in either building via optical frequency combs. 
Thus, either optical or microwave signals have to be exchanged between the buildings. The first is preferable as possible uncertainties, e.g., due to temperature induced phase variations on the microwave cable, are minimized with shorter leads inside a single building. All frequencies are measured by dead time-free counters in phase averaging mode with a gate time of one second.

\subsection{Sr lattice clock} The system has been described in detail previously \cite{fal11,fal14,alm15,gre16}. 
Compared to the most recent uncertainty evaluation \cite{gre16}, two uncertainty contributions have been further reduced.
The treatment of the lattice light shift has been changed and now employs an `operational magic wavelength' \cite{kat15}, reducing this contribution by a factor of 3. 
Secondly, we have more precisely taken into account the shift from blackbody radiation (BBR) emitted by the strontium oven on the atoms. 
For this characterization, a mechanical shutter located close to the oven was used to block radiation emitted into the vacuum chamber in one of two interleaved stabilizations of the interrogation laser to the atomic reference transition. 
These measurements allow for a much more accurate characterization of the light shift (by a factor of 8) than the previous geometry considerations \cite{fal14}. 
The current uncertainty budget is shown in Tab.~\ref{tab:srunc}. 
For most of the here presented absolute frequency measurement intervals, the total uncertainty was higher but always remained below $3 \times 10^{-17}$.
\begin{table}
	\caption{\label{tab:srunc}Uncertainty budget of PTB's laboratory Sr lattice clock. Corrections and uncertainties are given in fractional frequency units.}	
	\begin{ruledtabular}
	\begin{tabular}{ldd}
		\textrm{Effect} 	& \multicolumn{1}{c}{\textrm{Correction}} & \multicolumn{1}{c}{\textrm{Uncertainty } $u_{b,{\rm Sr}}$ }\\
		&	\multicolumn{1}{c}{$(10^{-17})$}	& \multicolumn{1}{c}{$(10^{-17})$} \\	
		\colrule
		Lattice light shift:&0.36&0.31\\
		BBR, ambient:&490.64&1.37\\
		BBR, oven:&0.30&0.12\\
		Second-order Zeeman:&13.41&0.10\\
		Collisions:&0.06&0.09\\
		Servo error:&0.00&0.07\\
		Tunnelling:&0.00&0.48\\
		DC Stark shift:&0.20&0.07\\
		Background gas collisions:&0.19&0.19\\
		Other:&0.00& <0.01\\
		\colrule
		Total:&	505.2&	1.5\\		
	\end{tabular}
	\end{ruledtabular}
\end{table}
\normalsize
\begin{figure}[tb]
	\includegraphics[width=0.45\textwidth]{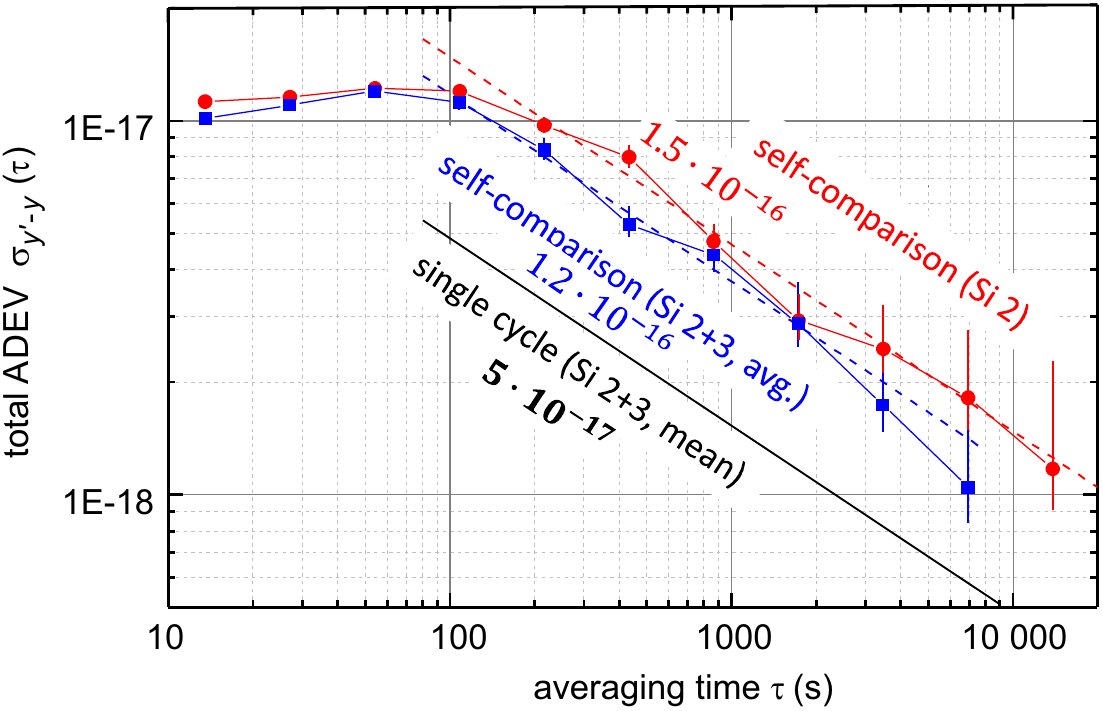}
	\caption{Instabilities of the differential servo signals of two independent, interleaved servos running on the PTB Sr lattice clock using one cryogenic silicon cavity (Si~2) and averaging the frequency of two cavities (Si~2 + 3). We can infer an instability of the Sr frequency standard of $5\times 10^{-17} / \sqrt{\tau/\mathrm{s}}$ for a single stabilization (black line).}
	\label{fig:srinstab}
\end{figure}

The frequency instability of the Sr lattice clock has been improved by transferring the ultralow instability of a laser stabilized to a cryogenic Si cavity \cite{mat17a} to the 429~THz interrogation laser \cite{hae15a} of the Sr clock via an optical frequency comb \cite{hag13,ben19}. 
The fractional frequency difference $y - y'$ between two interleaved stabilizations on the lattice clock exhibits a total Allan deviation (ADEV) of $\sigma_{y-y'}=1.5 \times 10^{-16}/\sqrt{\tau/\mathrm{s}}$ (Fig.~\ref{fig:srinstab}). 
As we had access to two cryogenic cavities for some time, we were able to further reduce the laser noise by averaging the laser frequencies of two cavity-stabilized lasers by rf-techniques in the stability transfer.
The clock instability was thus reduced to $\sigma_{y-y'}=1.2 \times 10^{-16}/\sqrt{\tau/\mathrm{s}}$. 
If the full measurement time is dedicated to a single stabilization we infer a fractional instability of the Sr frequency standard of $\sigma_{y, {\rm Sr}}=5 \times 10^{-17}/\sqrt{\tau/\mathrm{s}}$ by applying the noise model presented in \cite{alm15}. 
This result is on par with the currently best instabilities of optical clocks \cite{oel19}. 

During frequency measurements, the Sr standard is usually operated including a dead time of 1~s to reduce the thermal load on the vacuum chamber caused by power dissipation in magnetic field coils. This increases the clock's instability to $\sigma_{y, {\rm Sr}} = 1 - 2 \times 10^{-16}/\sqrt{\tau/\mathrm{s}}$, which is still well below the maser instability, but reduces the blackbody radiation uncertainty that dominates the systematic uncertainty $u_{b,{\rm Sr}}$ (Tab.~\ref{tab:srunc}).

\subsection{Cs primary frequency standards}
From 2017 to 2019, both caesium fountains CSF1 and CSF2 provided numerous calibrations of the scale unit of TAI. For this purpose the fountains were running in primary frequency standard (PFS) mode \cite{wey18}, i.e.,  at the highest available level of accuracy, for several weeks. For the $^{87}$Sr clock transition frequency measurements at hand, only data from such intervals have been evaluated. The comparison with TAI calibration data  \cite{Circular_T_Cs} from other primary and secondary frequency standards substantiates the performance of both fountains 
during the evaluation intervals.
\begin{figure}[tb]
	\includegraphics[width=0.45\textwidth]{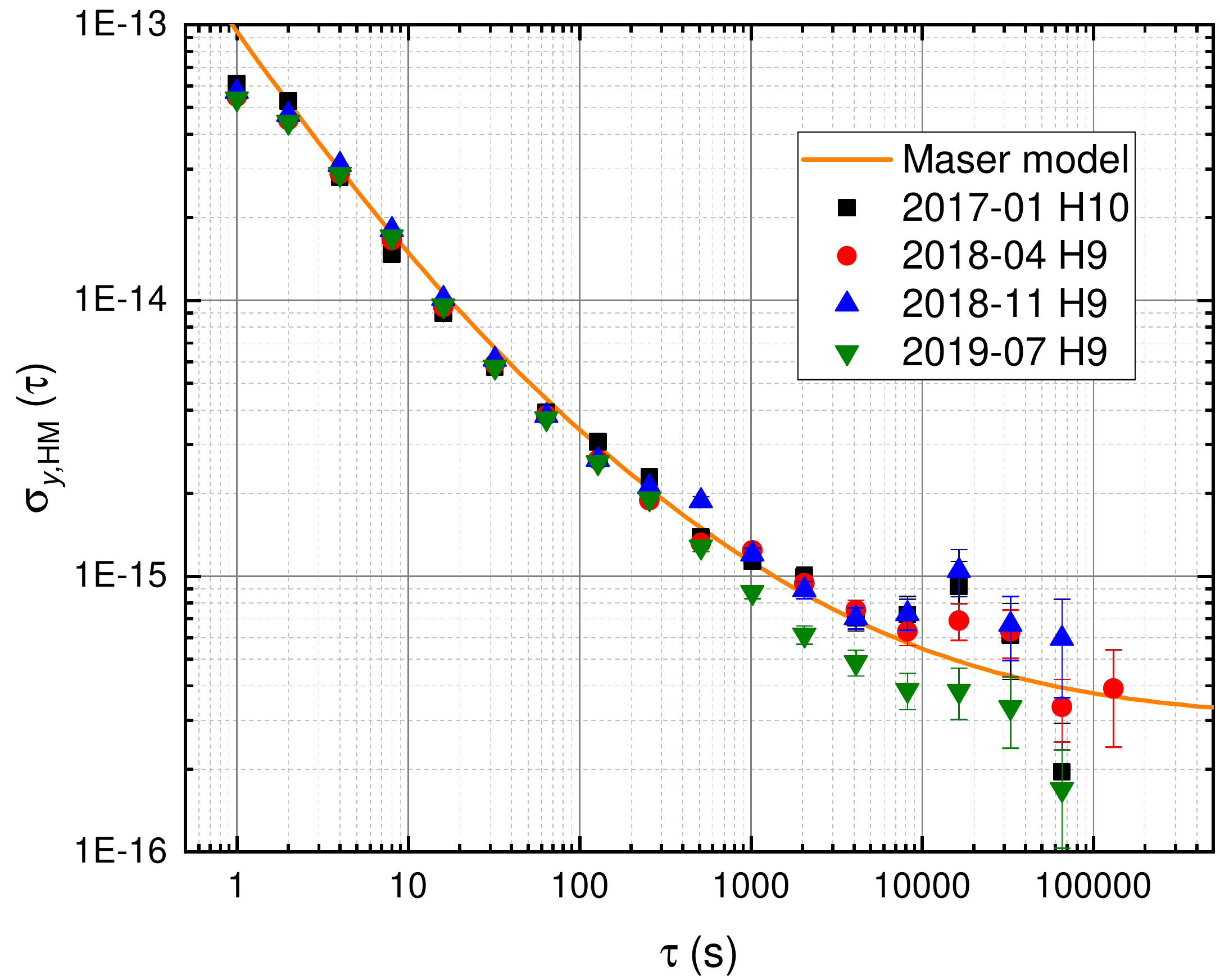}
	\caption{Instabilities of the hydrogen masers H9 and H10 as determined by comparisons against the Sr lattice clock. Linear drifts of less than $1.1\times 10^{-16}$ per day were	 removed from the data.}
	\label{fig:HM}
\end{figure}

In contrast to the typical PFS mode evaluations of CSF1 \cite{wey18}, we decided  for the evaluations of the comparably short optical frequency measurement intervals to disentangle the statistical and systematic parts of the collisional shift evaluations by using collisional shift coefficients, which were obtained just in the actual measurement time. Like in CSF2, we include the dominating statistical uncertainty of the individual collisional shift evaluations in the overall statistical uncertainty of the respective measurement intervals. Here, this treatment is advantageous, since the dominating type A part of the collisional shift uncertainty is reduced by the large number of short measurements.  At the same time, the removal of the type A part of the collisional shift uncertainty  from the overall type B uncertainty results in lower systematic uncertainties
(Tab.~\ref{tab:data}) compared to the typical PFS mode evaluations of CSF1 
\cite{wey18}. The systematic uncertainties of CSF2 
correspond to the previously published value \cite{wey18}. Including the statistical uncertainty of the collisional shift evaluation, the effective ADEVs \cite{wey18} are $\sigma_{y,{\rm Cs}}=3.3 \times 10^{-13}/ \sqrt{\tau/\mathrm{s}}$ and $\sigma_{y,{\rm Cs}}=1.6 \times 10^{-13} / \sqrt{\tau/\mathrm{s}}$ for CSF1 and CSF2, respectively.

\subsection{Hydrogen maser}  
During the period from 2017 through 2019, either of the active hydrogen masers H9 and H10 of the type VREMYA-CH VCH-1003M provided the rf-reference at PTB. Their performance was constantly monitored by a cluster of masers. To evaluate the uncertainty introduced by the application of the maser as flywheel oscillator in Sec.~\ref{Sec:data}, a maser noise model is required. It is based on the instability of the frequency ratio of the Sr standard and the respective maser from four long measurement intervals in January 2017, April 2018, November 2018 and July 2019. The ADEV shown in Fig.~\ref{fig:HM} was calculated under consideration of the gaps in the datasets \cite{ses08} caused by the down times of the Sr clock or the optical frequency combs that are used to measure the frequency ratio. 
A linear frequency drift was removed from the data as the evaluation method in Sec.~\ref{Sec:data} is insensitive to it.

The fractional frequency instabilities shown in Fig.~\ref{fig:HM} are dominated by the maser noise as the optical frequency standard shows considerably better stability on all time scales. We model the single-sided power spectral density $S_y(f)$ of the maser by the power law expansion $S_{y} = \sum_{\alpha=-1}^1{h_\alpha\,f^\alpha}$ adding flicker frequency (FFN), white frequency (WFN) and flicker phase noise (FPN). The coefficients $h_\alpha$ are adjusted such that the calculated ADEV $\sigma_{y, \rm{HM}}$ \cite{daw07,ben15} matches the observations in Fig.~\ref{fig:HM}. We find the coefficients 
\begin{eqnarray}
		h_1 = 4.3 \times 10^{-26}\;\textrm{Hz}^{-2} &\Rightarrow& \; \sigma_{y,\rm{FPN}} = 7 \times 10^{-14} \; \mathrm{s}/\tau \nonumber \\
    h_0 = 1.2 \times 10^{-27}\;\textrm{Hz}^{-1} &\Rightarrow& \; \sigma_{y,\rm{WFN}} = 2.4 \times 10^{-14} \; \sqrt{\mathrm{s}/\tau} \nonumber \\
    h_{-1} = 6.5 \times 10^{-32} &\Rightarrow& \; \sigma_{y,\rm{FFN}} = 3 \times 10^{-16} \label{eq:HM}		
\end{eqnarray}
using a cut-off frequency of 0.5~Hz.

%
%
\section{Data analysis and results}
\label{Sec:data}
%

Data from the Sr lattice clock were mainly recorded for other purposes than absolute frequency measurements. Therefore, they contain a relatively large number of gaps. Hence, the measurement intervals are often too short to achieve satisfactory statistical uncertainty in a direct comparison against a primary Cs clock. Applying the method described in \cite{gre16}, the results can be improved considerably: Due to the low instabilities of the masers and of the optical clock, in a first step their frequency ratio is evaluated with low statistical uncertainty on the combined uptimes of duration $T_{\rm Sr}$ of the optical clock, frequency comb, and links. We perform the conversion from optical to microwave domain preferentially on the frequency comb located in the same building as the microwave standards (Sec.~\ref{Sec:clocks}). For intervals, in which the Sr lattice clock was running, but neither of the optical links between the buildings was operational, we complemented the dataset by ratios measured via the microwave cable between the buildings, i.e. performing the conversion from microwave to optical domain at the frequency comb in the building with the lattice clock.

In a second step, the maser frequency is evaluated on a longer and uninterrupted time window of duration $T_{\rm Cs}$ by the primary frequency standards to lower the statistical uncertainty $u_{a,{\rm Cs}} = \sigma_{y,{\rm Cs}}(T_{\rm Cs})$ related to the fountain clock. Combining both measurements gives an absolute frequency measurement of the Sr lattice clock against the caesium fountains with a statistical uncertainty lower than for a direct comparison without the maser as flywheel. However, an extrapolation uncertainty $u_{\rm ext}$ must be added  due to the different averaging times of both measurements.

The extrapolation uncertainty can be calculated analytically if the single-sided power spectral density $S_y(f)$ of the flywheel oscillator is well characterized (Sec.~\ref{Sec:clocks}). The approach relies on Parseval's theorem and uses a weighting function $g(t)$ that represents the uptimes of the Sr and Cs clocks. $g(t)$ is defined as the difference of two weighting functions $g_i(t)$ with $i \in \left\{ {\rm Sr, \; Cs} \right\}$. $g_i(t)$ is $1/T_i$ if the respective clock is operated at $t$ and zero otherwise. Here, $T_i$ is the total operation time of clock $i$ in seconds, hence $\int g_i(t) dt = 1$. We find that \cite{gre16}
\begin{equation}
u_{\rm ext}^2 = \int ^\infty _0 S_y(f) \left| G(t) \right|^2 df
\label{eq:u_ext}
\end{equation}
with $G(t) =  \int g(t) \exp(-2\pi {\rm i} f t) dt$. Note that this approach is insensitive to linear drifts of the flywheel's frequency if the centers of gravity of both $g_i$ coincide. Otherwise,  drift correction has to be applied. Here, the centers of gravity deviate for some measurement intervals by few hours. As the frequency drift of the masers is below $1.1\times 10^{-16}$ per day, the correction and its uncertainty contribution are negligible.
 
\begin{figure}[tb]
	\includegraphics[width=0.45\textwidth]{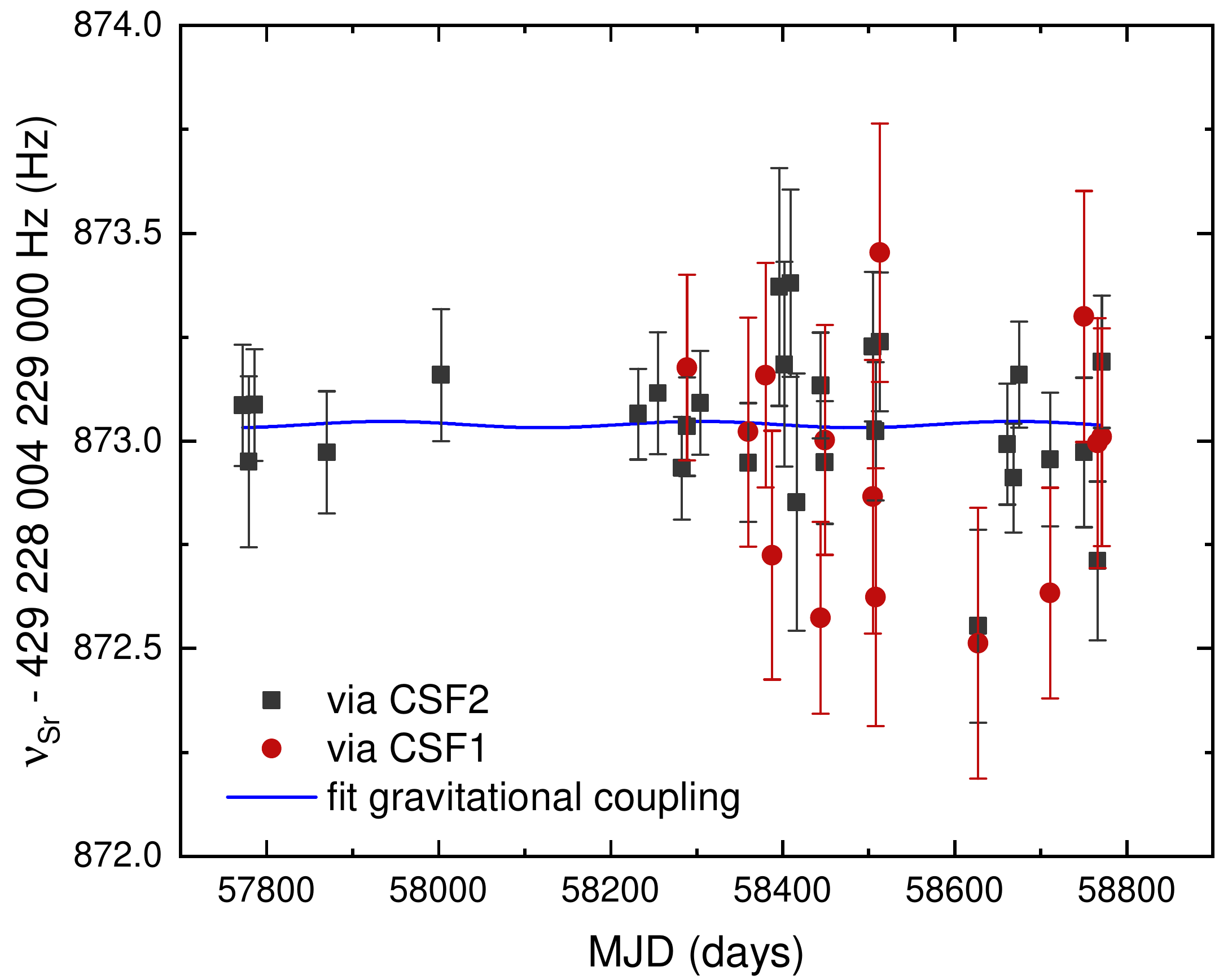}
	\caption{Absolute frequencies $\nu_{\rm Sr}$ of the (5s$^2$)~$^1$S$_0$--(5s5p)~$^3$P$_0$ transition in $^{87}$Sr measured by the Cs fountain clocks CSF1 and CSF2. Modified Julian dates (MJD), frequencies, and uncertainties $u$ from Tab.~\ref{tab:data}. The curve shows the fit to obtain limits on the coupling $\mu$ to the gravitational potential of the Sun (Sec.~\ref{Sec:mu}).}
	\label{fig:SrCSF}
\end{figure}

As discussed in \cite{gre16}, the statistical uncertainty of such a measurement is $u_a^2 = u^2_{\rm ext} + \sigma_{y, {\rm Sr}}^2(T_{\rm Sr}) + \sigma^2_{y, {\rm Cs}}(T_{\rm Cs})$, where the second term is negligibly small in our measurements. We have investigated 
at which $T_{\rm Cs}$ the combined statistical uncertainty $u_a$ is minimized for each measurement window. The length of filled gaps was  restricted to a maximum of one day. We find that for any individual measurement the value of $u_a$ is only weakly dependent on the choice of $T_{\rm Cs}$. 
Information on measurement intervals including dataset lengths, uncertainties, and the measured frequencies of the Sr clock are summarized in Tab.~\ref{tab:data}. Figure~\ref{fig:SrCSF} shows the measured absolute frequencies $\nu_{\rm Sr}$.

The uncertainties $u$ of the transition frequencies $\nu_{\rm Sr}$ in Tab.~\ref{tab:data} are dominated by the statistical uncertainty $u_{a,{\rm Cs}}$. Like the small statistical uncertainty contribution from the Sr frequency standard and $u_{\rm ext}$ (to a very large extent), it is	 uncorrelated between the measurement windows. 
As the systematic uncertainties $u_b$ of the frequency standards, in particular of CSF1, vary throughout the dataset for each fountain, weighted averages taking into account correlated ($u_b$) and uncorrelated ($u_a$, $u_{\rm ext}$) uncertainty contributions are calculated \cite{cow98}. We find the frequency values and uncertainties given in the last line of Tab.~\ref{tab:data}. The uncertainties are $2.4 \times 10^{-16}$ and $1.9 \times 10^{-16}$ in fractional frequency units for CSF1 and CSF2, respectively. Considering the systematic uncertainties of both fountain clocks as uncorrelated and neglecting the small correlations introduced by $u_{\rm ext}$ for windows with measurements by both fountains, we calculate a weighted average frequency of $429\:228\:004\:229\:873.00(0.07)$~Hz. The fractional frequency uncertainty of $1.5 \times 10^{-16}$ is even slightly below the record one reached for an Yb lattice clock \cite{mcg19}. These results are in very good agreement with the numerous previous determinations of $\nu_{\rm Sr}$ by a variety of groups \cite{lud06,let06a,tak06,boy07,bai08,cam08b,hon09,fal11,let13,gre16,fal14,yam12,mat12,aka14,tan15,lin15,hob16a,hac17,hac17a} (Fig.~\ref{fig:mudot}) and the value for $\nu_{\rm Sr}$ recommended as secondary representation of the second \cite{BIPM18}.

\begin{table*}
	\caption{\label{tab:data}Summary of the relevant information on the absolute frequency measurements for the respective measurement intervals labelled by the modified Julian date (MJD). $\Delta\nu$ is the value to be added to $429\,228\,004\,228\,000$~Hz to get the frequency $\nu_{\rm Sr}$ of the (5s$^2$)~$^1$S$_0$--(5s5p)~$^3$P$_0$ transition in $^{87}$Sr measured by the fountain clocks CSF1 and CSF2, respectively. The statistical uncertainty contribution from the Sr frequency standard is below $1 \times 10^{-18}$ for each measurement and therefore negligible. The total operation time of the Sr frequency standard is 59 days.}	
	\begin{ruledtabular}
	\begin{tabular}{l||rll||lll|ll||lll|ll}
					\multicolumn{4}{c||}{}			& \multicolumn{5}{c||}{CSF1}	& \multicolumn{5}{c}{CSF2} \\
		\colrule
			MJD		& $T_{\rm Sr}$& $u_{b,{\rm Sr}}$& $u_{\rm ext}$& $T_{\rm Cs}$& $u_{a,{\rm Cs}}$& $u_{b,{\rm Cs}}$ &  $\Delta\nu$ & $u$  &$T_{\rm Cs}$& $u_{a,{\rm Cs}}$& $u_{b,{\rm Cs}}$ & $\Delta\nu$  &  $u$ \\
	(days) 		&	(s)					& \multicolumn{2}{c||}{($10^{-16}$)} & (days)				 & 	\multicolumn{2}{c|}{($10^{-16}$)} & (Hz)& (Hz)&(days)				 & 	\multicolumn{2}{c|}{($10^{-16}$)}        & (Hz)& (Hz)\\	
		\colrule
57772	&	197\,872	&	0.29	&	0.9	&		&		&		&		&		&	3.99	&	2.8	&	1.7	&	873.09	&	0.15	\\
57779	&	186\,418	&	0.28	&	3.0	&		&		&		&		&		&	2.13	&	3.3	&	1.7	&	872.95	&	0.21	\\
57786	&	208\,520	&	0.2	&	1.2	&		&		&		&		&		&	4.67	&	2.3	&	1.7	&	873.09	&	0.13	\\
57870	&	143\,640	&	0.25	&	1.4	&		&		&		&		&		&	4.11	&	2.6	&	1.7	&	872.97	&	0.15	\\
58003	&	153\,004	&	0.18	&	1.4	&		&		&		&		&		&	3.71	&	2.9	&	1.7	&	873.16	&	0.16	\\
58232	&	377\,157	&	0.2	&	0.8	&		&		&		&		&		&	7.66	&	1.7	&	1.7	&	873.06	&	0.11	\\
58255	&	98\,997	&	0.16	&	1.8	&		&		&		&		&		&	3.68	&	2.3	&	1.8	&	873.12	&	0.15	\\
58283	&	230\,162	&	0.17	&	1.1	&		&		&		&		&		&	5.33	&	2.0	&	1.8	&	872.93	&	0.12	\\
58289	&	260\,066	&	0.17	&	1.0	&	5.44	&	4.9	&	1.4	&	873.18	&	0.22	&	5.44	&	1.9	&	1.8	&	873.03	&	0.12	\\
58304	&	194\,721	&	0.17	&	1.2	&		&		&		&		&		&	5.21	&	1.9	&	1.8	&	873.09	&	0.12	\\
58360	&	277\,031	&	0.15	&	1.4	&	3.57	&	6.0	&	1.9	&	873.02	&	0.28	&	3.57	&	2.5	&	1.8	&	872.95	&	0.14	\\
58380	&	67\,635	&	0.2	&	2.3	&	3.62	&	5.6	&	1.7	&	873.16	&	0.27	&		&		&		&		&		\\
58388	&	99\,503	&	0.25	&	1.9	&	2.88	&	6.6	&	1.4	&	872.72	&	0.30	&		&		&		&		&		\\
58396	&	107\,452	&	0.21	&	1.7	&		&		&		&		&		&	3.55	&	6.2	&	1.7	&	873.37	&	0.29	\\
58402	&	266\,345	&	0.19	&	0.9	&		&		&		&		&		&	4.56	&	5.4	&	1.7	&	873.18	&	0.25	\\
58409	&	336\,176	&	0.16	&	0.8	&		&		&		&		&		&	5.11	&	4.9	&	1.7	&	873.38	&	0.22	\\
58416	&	67\,507	&	0.19	&	2.2	&		&		&		&		&		&	2.71	&	6.6	&	1.7	&	872.85	&	0.31	\\
58444	&	289\,470	&	0.2	&	0.9	&	5.67	&	5.1	&	1.5	&	872.57	&	0.23	&	5.54	&	2.2	&	1.7	&	873.13	&	0.13	\\
58449	&	187\,058	&	0.16	&	1.3	&	3.67	&	6.2	&	1.4	&	873.00	&	0.28	&	3.67	&	2.7	&	1.7	&	872.95	&	0.15	\\
58505	&	68\,058	&	0.18	&	2.2	&	2.78	&	7.2	&	1.4	&	872.87	&	0.33	&	2.78	&	3.1	&	1.7	&	873.23	&	0.18	\\
58508	&	102\,630	&	0.2	&	2.1	&	3.20	&	6.8	&	1.3	&	872.62	&	0.31	&	3.20	&	2.7	&	1.7	&	873.02	&	0.17	\\
58513	&	110\,455	&	0.19	&	1.9	&	3.49	&	6.7	&	1.9	&	873.45	&	0.31	&	3.49	&	2.9	&	1.7	&	873.24	&	0.17	\\
58627	&	85\,846	&	0.18	&	2.3	&	3.00	&	7.1	&	1.3	&	872.51	&	0.33	&	3.00	&	4.6	&	1.7	&	872.55	&	0.23	\\
58661	&	177\,461	&	0.16	&	1.4	&		&		&		&		&		&	4.93	&	2.6	&	1.7	&	872.99	&	0.15	\\
58668	&	225\,900	&	0.15	&	1.1	&		&		&		&		&		&	5.33	&	2.2	&	1.7	&	872.91	&	0.13	\\
58675	&	265\,248	&	0.19	&	1.0	&		&		&		&		&		&	5.67	&	2.2	&	1.7	&	873.16	&	0.13	\\
58711	&	48\,496	&	0.16	&	2.2	&	3.80	&	5.2	&	1.7	&	872.63	&	0.25	&	3.80	&	2.5	&	1.7	&	872.96	&	0.16	\\
58750	&	59\,726	&	0.17	&	2.2	&	3.17	&	6.3	&	2.3	&	873.30	&	0.30	&	3.17	&	3.1	&	1.7	&	872.97	&	0.18	\\
58766	&	97\,600	&	0.18	&	2.1	&	3.31	&	6.4	&	1.9	&	872.99	&	0.30	&	3.31	&	3.5	&	1.7	&	872.71	&	0.19	\\
58771	&	126\,282	&	0.16	&	1.6	&	3.56	&	5.7	&	1.6	&	873.01	&	0.26	&	3.56	&	2.9	&	1.7	&	873.19	&	0.16	\\
		\colrule
\textbf{Average} 		&					&				&			&			&				&				& \textbf{872.92}  &	\textbf{0.11} 	&				&			&			&	\textbf{873.05}		&	\textbf{0.08}	\\		
	\end{tabular}
	\end{ruledtabular}
\end{table*}
\normalsize

%
\section{Gravitational coupling and drift of $m_\mathrm{p}/m_\mathrm{e}$}
\label{Sec:mu}
%

Extended datasets of accurate clock comparisons have been analyzed to test our understanding of physics. Typically, a potential violation of the Einstein equivalence principle (EEP) is investigated that would manifest itself as drift or modulation of the measured frequency ratio. A violation may be caused by temporal variations of fundamental constants or their coupling to a gravitational potential \cite{let13,hun14,mcg19,god14,dzu17,ros08,gue12,pei13,ash18}. We first investigate the data in Tab.~\ref{tab:data} with respect to a coupling of the proton--electron mass ratio $\mu$ to the gravitational potential of the Sun. In the second part of this section, we address a potential temporal variation $\dot{\mu}$ on an extended dataset. 

A coupling of the Sr/Cs frequency ratio to the Sun's gravitational potential would cause a modulation of $\nu_{\rm Sr}$ of  
\begin{equation}
\nu_{\rm Sr}(t) = \nu_0 \left(1+A\cos\left[ 2\pi(t-t_0)/T_\mathrm{0}  \right] \right)
\end{equation}	
with $T_\mathrm{0} \approx 365.260$~days (the duration of the anomalistic year). 
For this analysis, $t_0$ is the perihelion of 2018. 
Fitting this expression to the data, we find $A= -18(86)\times 10^{-18}$. 
We treated the uncertainties of all $\Delta\nu$ in Tab.~\ref{tab:data} as uncorrelated. 
Although this approach neglects correlations introduced by $u_b$, for instance, this only causes an overestimating of the uncertainty of the fit parameter $A$. 
The reduced $\chi^2$ of 1.3 of the fit indicates that the fit is statistically well behaved.
With annual variation of the gravitational potential $\Delta \Phi = (\Phi_{\rm max} -\Phi_{\rm min})/2 \approx 1.65 \times 10^{-10} \cdot {\rm c}^2$ (c being the speed of light in vacuum), we find the coupling coefficient
\begin{equation}
\beta_{\rm Sr,Cs} = \frac{A}{\Delta \Phi / c^2} = -1.1(5.2)\times 10^{-7},
\label{eq:beta}
\end{equation}
thus no violation of the EEP is found. The limit set by our data is about a factor of two less stringent than the ones set by Ashby \emph{et al.} \cite{ash18} or Dzuba and Flambaum \cite{dzu17}. In principle, our data could be supplemented as in \cite{mcg19} by measurements performed by other groups, but the exact measurement intervals and their individual frequency values of past measurements are often not reported, which is why we refrain from this approach.

Effects causing a violation of EEP could be variations of the fine-structure constant $\alpha$, of  the ratio of light quark mass to the quantum chromodynamics (QCD) scale $X_q = m_q / \Lambda_{\rm QCD}$, or of $\mu$. 
For each effect and pair of compared clocks $X,Y$, sensitivity coefficients 
$\Delta K_\epsilon(X,Y) = d\ln(\nu_X/\nu_Y)/d\ln(\epsilon)$ with 
$\epsilon \in \left\{ \alpha, X_q, \mu \right\}$ 
have been determined through atomic and nuclear structure calculations \cite{fla06b, din09}.
For the pair (Sr,Cs) these calculations yield 
$\Delta K_\alpha({\rm Sr, Cs}) = -2.77$, 
$\Delta K_{X_q}({\rm Sr, Cs}) = -0.002$, 
and $\Delta K_\mu({\rm Sr, Cs}) = 1$ . 
Thus the coupling to the gravitational potential can be expressed as 
$\beta_{\rm Sr,Cs} = \sum_\epsilon \Delta K_\epsilon({\rm Sr,Cs})\,k_\epsilon$ 
with species-independent coupling coefficients 
$k_\epsilon=d\ln(\epsilon)/d(\Phi/c^2)$ 
for the respective effects.

Using limits on the other parameters' variations from complementary investigations as in \cite{mcg19}, 
$k_\alpha = 0.54(1.02) \times 10^{-7}$ \cite{dzu17} from the Al$^+$/Hg$^+$ frequency ratio which is only sensitive to variations of $\alpha$,
and $k_{X_q} = -2.6(2.6) \times 10^{-6}$ from the H,Cs pair \cite{ash18}, 
we derive a refined coupling parameter $k_\mu$ from our $\beta_{\rm Sr,Cs}$ (Eq.~\ref{eq:beta}): 
\begin{equation}
k_\mu = 3.5(5.9)\times 10^{-7}.
\label{eq:k_mu}
\end{equation}
This result is a factor of 1.6 more stringent than the result by McGrew \emph{et al.} \cite{mcg19} for the combined data of Yb and Sr lattice clocks.
\begin{figure}[tb]
	\includegraphics[width=0.45\textwidth]{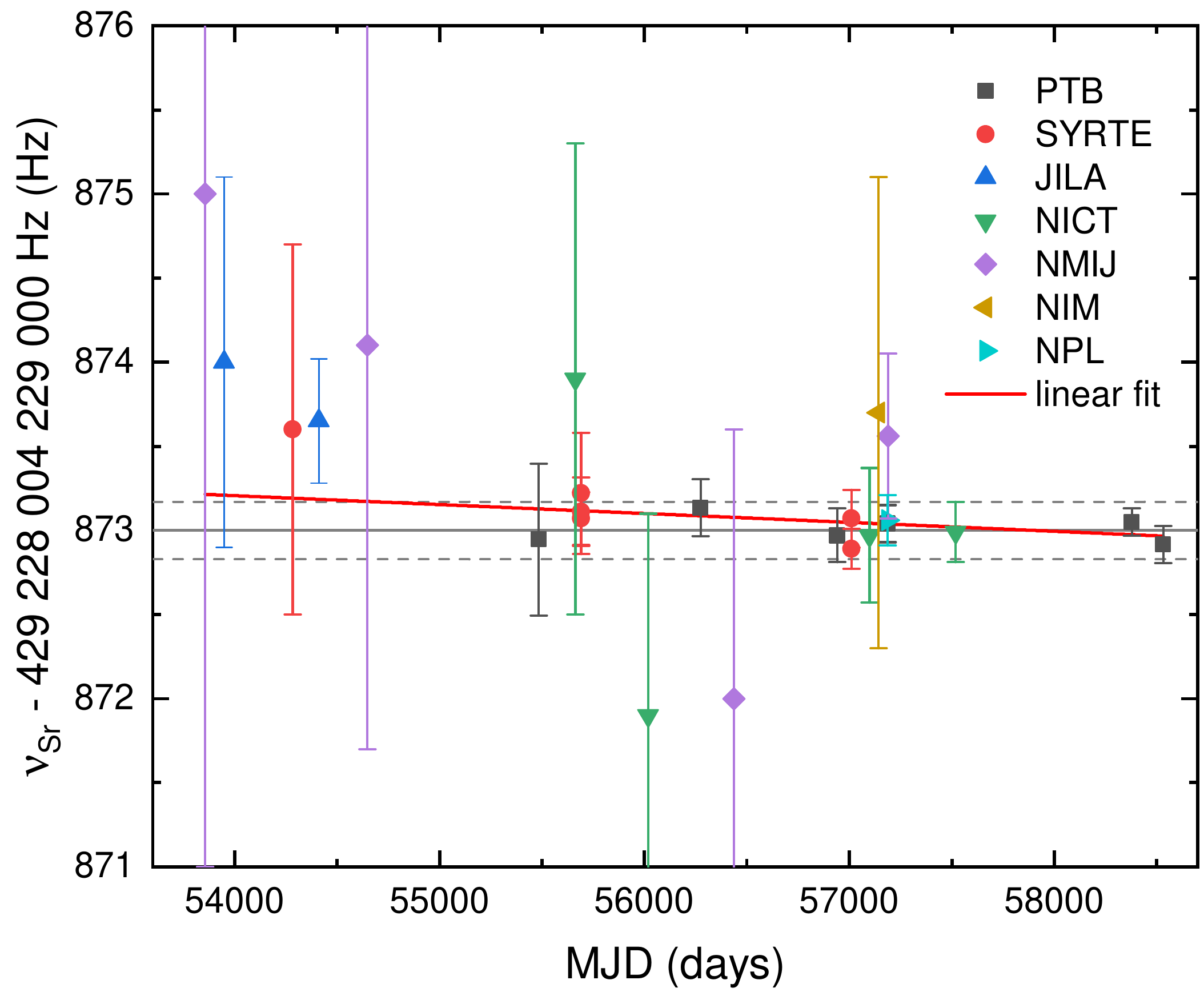}
	\caption{Overview of the measured values of the $^{87}$Sr 429~THz transition. The two squares on the right show the average values from Tab.~\ref{tab:data}. The horizontal gray line shows the value of the frequency recommended as secondary representation of the second (dashed lines: uncertainty). The red line is a fit to derive $\dot{\mu}/\mu$. }
	\label{fig:mudot}
\end{figure}

To analyze our data for temporal variation $\dot{\mu}/\mu$ we combine them with previously published values of $\nu_{\rm Sr}$ 
\cite{tak06,boy07,bai08,cam08b,hon09,fal11,yam12,mat12,let13,fal14,aka14,tan15,lin15,hob16a,lod16,gre16,hac17,hac17a}, spanning in total more than 12 years. 
We use the averaged frequency values only and estimate, where necessary, the time of measurement, whose exact knowledge is of smaller relevance for this analysis than for the one above. Where averaged frequency values from measurements against different fountain clocks are available in a publication, we use those instead of the combined averages. Fitting a linear temporal variation of $\nu_{\rm Sr}$ (Fig.~\ref{fig:mudot}) yields a slope of $\dot{y} = -4.2(3.3) \times 10^{-17}$/a in fractional frequency units.
This result slightly improves on the limit given in \cite{mcg19} of $-4.9(3.6) \times 10^{-17}$/a that is based on Yb lattice clock data in addition to the same published Sr frequency measurements.

Expressing the variation $\dot{y} = \sum_i \Delta K_\epsilon(X,Y)\,d\ln(\epsilon)/dt$ as sum of three varying fundamental constants $\epsilon \in \left\{ \alpha, X_q, \mu \right\}$, and restricting two of them by published values
$\dot{X_q}/X_q = 0.14(9) \times 10^{-16}/\mathrm{a}$ from \cite{gue12} and 
the averaged value from \cite{god14,hun14} $\dot{\alpha}/\alpha = -1.4(2.0) \times 10^{-17}/\mathrm{a}$ (as both results are probably strongly correlated, we have kept the slightly lower uncertainty from \cite{hun14}), we obtain
\begin{equation}
\dot{\mu}/\mu = -6.9(6.5) \times 10^{-17}/\mathrm{a}.
\label{eq:mudot}
\end{equation}

This value confirms the rate $\dot{\mu}/\mu = -5.2(6.5) \times 10^{-17}$/a found in \cite{mcg19} obtained from Sr and Yb lattice clock frequency measurements. It is interesting to note that the uncertainty of both results is limited by that of $\dot{\alpha}/\alpha$ \cite{god14,hun14}.

%
\section{Conclusion}
\label{Sec:conc}
%

We have presented more than 40 measurements of the $^{87}$Sr 429~THz clock transition against PTB's primary fountain clocks CSF1 and CSF2. The uncertainty of the average frequency is 65~mHz ($1.5 \times 10^{-16}$ in fractional units), which is more accurate than the best absolute frequency measurements so far \cite{let13, hun14, gre16, lod16, mcg19, piz20}. Our result is a valuable contribution to the validation of $^{87}$Sr lattice clocks as secondary representation of the second as well as to a future redefinition of the SI unit second.

We have presented an updated uncertainty budget of PTB's Sr lattice clock at a fractional uncertainty level of $1.5 \times 10^{-17}$. Self-comparisons of two clock stabilization cycles run on the same apparatus indicate that our lattice clock can reach an instability of $\sigma_{y, {\rm Sr}}=5 \times 10^{-17}/\sqrt{\tau/\mathrm{s}}$, a level similar to the one reached in \cite{oel19}.

Using the frequency comparisons between Cs and Sr clocks, we have performed a test of the Einstein equivalence principle that -- in combination with data from \cite{dzu17,ash18} -- allows us to put more a stringent limit on the coupling constant between the proton--electron mass ratio $\mu$ and a gravitational potential of $k_\mu = 3.5(5.9)\times 10^{-7}$. The limit we set on $\dot{\mu}/\mu = -6.9(6.5) \times 10^{-17}/$a confirms the result given in \cite{mcg19}.

\begin{acknowledgments}
We acknowledge support by the project 18SIB05 ROCIT, which has received funding from the EMPIR programme co-financed by the Participating States and from the European Union’s Horizon 2020 research and innovation programme, and by the Deutsche Forschungsgemeinschaft (DFG, German Research Foundation) under Germany’s Excellence Strategy -- EXC-2123 QuantumFrontiers -- 390837967 and CRC~1228 DQ-\emph{mat} within project B02.
This work was partially supported by the Max Planck--RIKEN--PTB Center for Time, Constants and Fundamental Symmetries.
We acknowledge the fruitful collaboration with Jun Ye's group at JILA, Boulder in the development of cryogenic silicon resonators.
\end{acknowledgments}

\section*{References}



\end{document}